\documentclass[11pt,a4paper]{article}

\usepackage{bm,subfigure,jcappub}

\title{Flavor ratios of astrophysical neutrinos interacting with stochastic gravitational waves having arbitrary spectra}

\author{Maxim Dvornikov}
\affiliation{Pushkov Institute of Terrestrial Magnetism, Ionosphere and Radiowave Propagation (IZMIRAN), 108840 Troitsk, Moscow, Russia}

\emailAdd{maxdvo@izmiran.ru}

\abstract{
We study the evolution and oscillations of fixed massive neutrinos
interacting with stochastic gravitational waves (GWs). The energy
spectrum of these GWs is Gaussian, with the correlator of the amplitudes
being arbitrary. We derive the equation for the density matrix for flavor
neutrinos in this case. In the two flavors approximation, this equation
can be solved analytically. We find the numerical solution for the
density matrix in the general case of three neutrino flavors. We consider
merging binary black holes as sources of stochastic GWs with realistic
spectra. Both normal and inverted mass orderings are analyzed. We
discuss the relaxation of the neutrino fluxes in stochastic GWs emitted
mainly by supermassive black holes. In this situation, we obtain the
range of energies and the propagation lengths for which the relaxation
process is the most efficient. We discuss the application of our results
for the observation of fluxes of astrophysical neutrinos.
}


\begin{document}

\maketitle

\section{Introduction}

Flavor transformations of a neutrino beam, called neutrino flavor
oscillations, recently confirmed experimentally (see, e.g. ref.~\cite{Ago20}),
are the direct indication to the nonzero masses of these particles
and the mixing between different neutrino states. Various external fields
are known to contribute, or even enhance, neutrino oscillations~\cite{FukYan03}.
The gravitational interaction, in spite of its weakness, is supposed
to modify the neutrino oscillations process.

Neutrino oscillations in gravitational fields were first considered
in ref.~\cite{AhlBur96} and, subsequently, in multiple papers. Many
of them are reviewed in ref.~\cite{LucPet20}. A gravitational field
is supposed to modify the phase of a massive neutrino wave function~\cite{For97}.
This phase obeys the Hamilton-Jacobi equation, which should be written
down in curved spacetime.

It is important to study neutrino oscillations in gravitational fields
considering not only static fields, but also time dependent gravitational
backgrounds like a gravitational wave (GW). It is inspired by the
recent GWs detection reported in ref.~\cite{Abb16}. Now multiple
sources of GWs, mainly as coalescing binary compact objects, are catagolized
in ref.~\cite{Abb18}. Sources of GWs are expected to emit significant
neutrino fluxes~\cite{Mes19}. Thus, emitted GWs can influence the
propagation and oscillations of astrophysical neutrinos. There are
active searches of neutrinos emitted by merging binaries, which are
the sources of GWs~\cite{Aar20b}.

We studied neutrino flavor and spin (i.e., transitions between active
left and sterile right states) oscillations under the influence of
gravitational fields in refs.~\cite{Dvo06,Dvo13,Dvo20,Dvo19b,Dvo19}.
Neutrino spin oscillations in GWs were considered in ref.~\cite{Dvo19b}.
The evolution of the spin of a fermion in GW was discussed in ref.~\cite{ObuSilTer17}.
The action of GWs on the propagation and flavor oscillations of astrophysical
neutrinos was studied in refs.~\cite{Dvo19,KouMet19}.

In the present work, we continue our research in ref.~\cite{Dvo19}.
We study the evolution of a beam of fixed massive neutrinos interacting
with stochastic GWs with an arbitrary energy spectrum. In section~\ref{sec:DENSMATR},
we derive the equation for the density matrix for flavor neutrinos
and solve it analytically in the two flavors approximation. The numerical
solution of this equation for the general situation of the three neutrino
flavors is obtained in section~\ref{sec:APPL}, where we study astrophysical
applications. We consider coalescing binary BHs as sources of stochastic
GWs. We obtain the maximal energy and the minimal propagation length,
at which the fluxes of flavor neutrinos reach their asymptotic values.
We discuss our results and consider their implication for the observation
of astrophysical neutrinos in section~\ref{sec:DISC}. The averaged fluxes of neutrinos involved in flavor oscillations in vacuum are obtained in appendix~\ref{sec:FLUXES}.

\section{Density matrix evolution accounting for an arbitrary spectrum of
GWs\label{sec:DENSMATR}}

We study the system of three flavor neutrinos $(\nu_{e},\nu_{\mu},\nu_{\tau})$
with nonzero mixing. One can diagonalize the mass matrix of such neutrinos
using the neutrino mass eigenstates $\psi_{a}$, $a=1,2,3$, which
have the definite masses $m_{a}$. Flavor and mass eigenstates are
related by the matrix transformation,
\begin{equation}\label{eq:nupsi}
  \nu=U\psi,
\end{equation}
where $U$ is the mixing matrix.

In the most general situation of three neutrino flavors $\nu=(\nu_{e},\nu_{\mu},\nu_{\tau})$,
the mixing matrix in eq.~(\ref{eq:nupsi}) can be parameterized in
the following form~\cite[pgs.~111\textendash 116]{GonMalSch16}:
\begin{equation}\label{eq:U3f}
  U=
  \left(
    \begin{array}{ccc}
      1 & 0 & 0\\
      0 & c_{23} & s_{23}\\
      0 & -s_{23} & c_{23}
    \end{array}
  \right)
  \cdot
  \left(
    \begin{array}{ccc}
      c_{13} & 0 & s_{13}e^{-\mathrm{i}\delta_{\mathrm{CP}}}\\
      0 & 1 & 0\\
      -s_{13}e^{\mathrm{i}\delta_{\mathrm{CP}}} & 0 & c_{13}
    \end{array}
  \right)
  \cdot
  \left(
    \begin{array}{ccc}
      c_{12} & s_{12} & 0\\
      -s_{12} & c_{12} & 0\\
      0 & 0 & 1
    \end{array}
  \right),
\end{equation}
where $c_{ab}=\cos\theta_{ab}$, $s_{ab}=\sin\theta_{ab}$, $\theta_{ab}$
are the corresponding vacuum mixing angles, and $\delta_{\mathrm{CP}}$
is the CP violating phase. The values of these parameters can be found
in ref.~\cite{Sal20}. The mixing matrix takes more simple form in
the frequently used two flavors approximation,
\begin{equation}
  U=
  \left(
    \begin{array}{cc}
      \cos\theta & \sin\theta\\
      -\sin\theta & \cos\theta
    \end{array}
  \right),
\end{equation}
where $\theta$ is the only mixing angle.

We suppose that these flavor neutrinos move in curved spacetime with
background gravitational field in the form of a plane gravitational
wave with the circular polarization propagating along the $z$-axis. The
interval in this case has the form~\cite{Buo07}, 
\begin{equation}\label{eq:metric}
  \mathrm{d}s^{2}=g_{\mu\nu}\mathrm{d}x^{\mu}\mathrm{d}x^{\nu}=  
  \mathrm{d}t^{2}-
  \left(
    1-h\cos\phi
  \right)
  \mathrm{d}x^{2}-
  \left(
    1+h\cos\phi
  \right)
  \mathrm{d}y^{2}+2\mathrm{d}x\mathrm{d}yh\sin\phi-\mathrm{d}z^{2},
\end{equation}
where $h$ is the dimensionless amplitude of the wave, $\phi=\left(\omega t-kz\right)$
is the phase of the wave, $\omega$ is frequency of the wave, and
$k$ is the wave vector. In eq.~(\ref{eq:metric}), we use the Cartesian
coordinates $x^{\mu}=(t,x,y,z)$.

The dynamics of flavor oscillations $\nu_{\alpha}\leftrightarrow\nu_{\beta}$
is described by the effective Schrodinger equation,
\begin{equation}\label{eq:effHam}
  \mathrm{i}\dot{\nu}=H_{f}\nu,
  \quad
  H_{f}=UH_{m}U^{\dagger},
  \quad
  H_{m}=H_{m}^{(\mathrm{vac})}+H_{m}^{(g)},
\end{equation}
where $H_{f}$ and $H_{m}$ are the effective Hamiltonians in the
flavor and mass bases, $H_{m}^{(\mathrm{vac})}=\tfrac{1}{2E}\text{diag}(m_{1}^{2},m_{2}^{2},m_{3}^{2})$
is the part of the Hamiltonian responsible for vacuum oscillations, $E\approx p$ is the mean neutrino energy, $p$ is the momentum of mass eigenstates.

Using the results of ref.~\cite{Pop06}, we have derived in ref.~\cite{Dvo19}
the contribution to $H_{m}$ from the neutrino interaction with GW.
We assume that 
\begin{equation}\label{eq:constr}
  \omega L|\beta_{a}-\beta_{b}|\ll1,
  \quad
  a,b=1,\dots,3,
\end{equation}
where $L$ is the neutrino propagation distance, $\beta_{a}=p/E_{a}$
is the velocity of the mass eigenstate, $p$ is the mean neutrino
momentum, and $E_{a}=\sqrt{m_{a}^{2}+p^{2}}$ is the energy of the
mass eigenstate. In this case, $H_{m}^{(g)}$ have only diagonal components
which have the form,
\begin{equation}\label{eq:gravHam}
  \left(
    H_{m}^{(g)}
  \right)_{aa}=
  -\frac{p^{2}h}{2E_{a}}\sin^{2}\vartheta\cos2\varphi\approx
  - hA
  \left(
    p - \frac{m_{a}^{2}}{2p}
  \right)
  \to
  hA\frac{m_{a}^{2}}{2E},
\end{equation}
where $\vartheta$ and $\varphi$ are the spherical angles fixing
the neutrino momentum with respect to the  wave vector $\mathbf{k} = (0,0,k)$ of GW and $A(\vartheta,\varphi)=\tfrac{1}{2}\sin^{2}\vartheta\cos2\varphi$. In eq.~(\ref{eq:gravHam}), we keep only the linear term in $h$.

In eq.~\eqref{eq:gravHam}, we assume that the momenta of all mass eigenstates are equal to $p$. Then, accounting for the fact that neutrinos are ultrarelativistic, i.e. $p\gg m_{a}$, we expand $H_{m}^{(g)}$ in the small parameters $m_a^2/p^2$ and omit the term which is equal for all mass eigenstates. It is known that such a term does not contribute to the dynamics of neutrino oscillations. Eventually, we replace $p \to E$ in the final expression. The same technique is used to derive the expression for $H_{m}^{(\mathrm{vac})}$ given above.

Finally, using eqs.~(\ref{eq:effHam}) and~(\ref{eq:gravHam}),
we get the total effective Hamiltonian for flavor eigenstates in the form,
\begin{equation}\label{eq:Hf}
  H_{f}=H_{0}+H_{1},
  \quad
  H_{0}=UH_{m}^{(\mathrm{vac})}U^{\dagger},
  \quad
  H_{1}=\xi H_{0},
\end{equation}
where $\xi=hA$. Note that $H_{0}$ in eq.~(\ref{eq:Hf}) is the
constant matrix.

Now we suppose that neutrinos interact with stochastic GWs, i.e. the
angles $\vartheta$ and $\varphi$, as well as $h$ are random functions
of time. In this case, it is convenient to study the evolution of
the density matrix $\rho$ rather than the wave function $\nu$. In
this approach, the diagonal elements of $\rho$ are the probabilities
to detect a certain flavor in a neutrino beam. Using the results of
ref.~\cite{LorBal94}, we get the equation for $\rho_{\mathrm{I}}(t)=\exp(\mathrm{i}H_{0}t)\rho(t)\exp(-\mathrm{i}H_{0}t)$
in the form,
\begin{equation}\label{eq:rhoI}
  \mathrm{i}\dot{\rho}_{\mathrm{I}}=[H_{\mathrm{I}},\rho_{\mathrm{I}}],
\end{equation}
where $H_{\mathrm{I}}=\exp(\mathrm{i}H_{0}t)H_{1}\exp(-\mathrm{i}H_{0}t)=H_{1}$.

The replacement of the density matrix $\rho\to\rho_{\mathrm{I}}$ means the consideration of the interaction picture~\cite{Wei96} for the neutrino flavor eigenstates. It is used frequently in quantum mechanics to develop a perturbative approach (see, e.g., ref.~\cite{LorBal94}).

The formal solution of eq.~(\ref{eq:rhoI}) can be represented in
the form of a series, which should be averaged over a certain time
interval. We suppose that the amplitudes of GW form the Gaussian stochastic
process. Thus, only even terms in this series survive since all odd
correlators, like $\left\langle h(t_{1})h(t_{2})h(t_{3})\right\rangle $
etc, are vanishing. Eventually one has
\begin{align}\label{eq:rhoIser}
  \left\langle
    \rho_{\mathrm{I}}
  \right\rangle (t)= & 
  \rho_{0}-
  \left\langle
    A^{2}
  \right\rangle
  [H_{0},[H_{0},\rho_{0}]]
  \int_{0}^{t}\mathrm{d}t_{1}
  \int_{0}^{t_{1}}\mathrm{d}t_{2}f(|t_{1}-t_{2}|)
  \nonumber
  \\
  & +
  \left\langle
    A^{2}
  \right\rangle^{2}
  [H_{0},[H_{0},[H_{0},[H_{0},\rho_{0}]]
  \int_{0}^{t}\mathrm{d}t_{1}
  \int_{0}^{t_{1}}\mathrm{d}t_{2}
  \int_{0}^{t_{2}}\mathrm{d}t_{3}
  \int_{0}^{t_{3}}\mathrm{d}t_{4}
  \nonumber \\
  & \times
  \big[
    f(|t_{1}-t_{2}|)f(|t_{3}-t_{4}|)+f(|t_{1}-t_{3}|)f(|t_{2}-t_{4}|)
    \notag
    \\
    & +
    f(|t_{1}-t_{4}|)f(|t_{2}-t_{3}|)
  \big]-
  \dotsc,
\end{align}
where $\rho_{0}=\rho_{\mathrm{I}}(0)=\rho(0)$ is the initial density
matrix, $f(|t_{1}-t_{2}|)=\left\langle h(t_{1})h(t_{2})\right\rangle $
is the correlator of GW amplitudes, and
\begin{equation}\label{eq:A2}
  \left\langle
    A^{2}
  \right\rangle =
  \frac{1}{2\pi^{2}}
  \int_{0}^{\pi}\mathrm{d}\vartheta\int_{0}^{2\pi}\mathrm{d}\varphi 
  A^{2}(\vartheta,\varphi)=
  \frac{3}{64},
\end{equation}
is the mean value of the angle factor squared.

To derive eq.(\ref{eq:rhoIser}) we assume that both $\vartheta$
and $\varphi$ have the $\delta$-correlated Gaussian distributions,
which is a reasonable assumption since stochastic GWs intersect a
neutrino trajectory randomly. However, unlike ref.~\cite{Dvo19},
we do not assume that the amplitude of GW has the same distribution,
i.e. $f(|t_{1}-t_{2}|)\not\sim\delta(t_{1}-t_{2})$ is the arbitrary
function.

After lengthy but straightforward calculations, we transform eq.~(\ref{eq:rhoIser})
to the form,
\begin{align}\label{eq:rhoIserfac}
  \left\langle
    \rho_{\mathrm{I}}
  \right\rangle (t) = &
  \rho_{0}-a
  \left\langle
    A^{2}
  \right\rangle
  [H_{0},[H_{0},\rho_{0}]]+\frac{1}{2!}
  \left(
    a
    \left\langle
      A^{2}
    \right\rangle
  \right)^{2}
  [H_{0},[H_{0},[H_{0},[H_{0},\rho_{0}]]-\dotsc,
\end{align}
where
\begin{equation}
  a(t)=\int_{0}^{t}\mathrm{d}t_{1}\int_{0}^{t_{1}}
  \mathrm{d}t_{2}f(|t_{1}-t_{2}|)=
  \int_{0}^{t}\mathrm{d}t_{1}(t-t_{1})f(|t_{1}|).
\end{equation}
One can check that eq.~(\ref{eq:rhoIserfac}) is the formal solution
of the following equation:
\begin{equation}\label{eq:rhoIeq}
  \frac{\mathrm{d}}{\mathrm{d}t}
  \left\langle
    \rho_{\mathrm{I}}
  \right\rangle (t)=
  -g(t)
  \left\langle
    A^{2}
  \right\rangle
  [H_{0},[H_{0},
  \left\langle
    \rho_{\mathrm{I}}
  \right\rangle
  (t)]],
\end{equation}
where 
\begin{equation}\label{eq:g}
  g(t)=\int_{0}^{t}\mathrm{d}t_{1}f(|t-t_{1}|),
\end{equation}
which is the generalization of the results of ref.~\cite{Dvo19}
for the arbitrary correlator $f(|t_{1}-t_{2}|)$ of the amplitudes
of GWs. If we choose $f(|t_{1}-t_{2}|)=2\tau\left\langle h^{2}\right\rangle \delta(t_{1}-t_{2})$,
where $\tau$ is the phenomenological correlation time, and use eqs.~(\ref{eq:A2})
and~(\ref{eq:g}), then eq.~(\ref{eq:rhoIeq}) reproduces eq.~(2.17)
in ref.~\cite{Dvo19}.

Let us study the two flavors approximation. In this case,
\begin{equation}
  H_{0}=\frac{\Delta m_{21}^{2}}{4E}(\bm{\sigma}\mathbf{n}),
  \quad
  \mathbf{n}=(\sin2\theta,0,-\cos2\theta),
\end{equation}
where $\Delta m_{21}^{2}=m_{2}^{2}-m_{1}^{2}$ and $\bm{\sigma}$ are the Pauli matrices. We can sum analytically
the series in eq.~(\ref{eq:rhoIserfac}). Indeed,
\begin{align}\label{eq:rhoI2fl}
  \left\langle
    \rho_{\mathrm{I}}
  \right\rangle (t)= &
  \rho_{0}-[\rho_{0}-
  (\bm{\sigma}\mathbf{n})\rho_{0}(\bm{\sigma}\mathbf{n})]\frac{\lambda}{2}
  \left(
    1-\frac{\lambda}{2!}+\frac{\lambda^{2}}{3!}-\dots
  \right)
  \nonumber
  \\
  & =
  \frac{1}{2}
  \left[
    \rho_{0}
    \left(
      1+e^{-\lambda}
    \right)+
    (\bm{\sigma}\mathbf{n})\rho_{0}(\bm{\sigma}\mathbf{n})
    \left(
      1-e^{-\lambda}
    \right)
  \right],
\end{align}
where $\lambda=a\left\langle A^{2}\right\rangle \left(\Delta m_{21}^{2}\right)^{2}/4E^{2}$.
If we choose the $\delta$-correlated Gaussian distribution, then
$a=\tau t\left\langle h^{2}\right\rangle $ and eq.~(\ref{eq:rhoI2fl})
reproduces the corresponding result of ref.~\cite{Dvo19}.

The correlation function $\left\langle h(t)h(0)\right\rangle =f(|t|)$
can be expressed in terms of the spectral density $S(f)$ as~\cite{KorKor68}
\begin{equation}\label{eq:corrS}
  \left\langle h(t)h(0)\right\rangle =\int_{0}^{\infty}\mathrm{d}f\cos(2\pi ft)S(f),
\end{equation}
where $f$ is the frequency measured in Hz. Instead of the spectral
density in eq.~(\ref{eq:corrS}) it is convenient to consider the
energy density of stochastic GWs, $\Omega(f)$, per logarithmic frequency
interval with respect to the closure density of the universe $\rho_{c}=\tfrac{3\mathrm{H}_{0}^{2}}{8\pi G}=0.53\times10^{-5}\text{GeV}\cdot\text{cm}^{-3}$~\cite{Chr19},
\begin{equation}
  S(f)=\frac{8G\rho_{c}}{\pi f^{3}}\Omega(f),
\end{equation}
where $\mathrm{H}_{0}$ is the Hubble constant and $G=6.9\times10^{-39}\,\text{GeV}^{-2}$
is the Newton constant. The function $g(t)$ in eq.~(\ref{eq:g})
takes the form,
\begin{equation}\label{eq:gspec}
  g(t)=\frac{4G\rho_{c}}{\pi^{2}}
  \int_{0}^{\infty}\frac{\mathrm{d}f}{f^{4}}\sin(2\pi ft)\Omega(f),
\end{equation}
which should be used in the differential eq.~(\ref{eq:rhoIeq}).

\section{Astrophysical applications\label{sec:APPL}}

In this section, we study flavor transformations of a neutrino beam
under the influence of a realistic stochastic GW background. This problem
was studied previously in ref.~\cite{Dvo19}, where coalescing black
holes (BHs) were considered. However, the correlator of GW amplitudes
was assumed in ref.~\cite{Dvo19} to be of the form, $\left\langle h(t_{1})h(t_{2})\right\rangle \sim\delta(t_{1}-t_{2})$.
Spectra of realistic stochastic GW backgrounds were mentioned in ref.~\cite{Abb19}
to be approximated by power laws, $\Omega(f)\sim f^{\alpha}$, with
the frequency $f$ being in a confined region. Thus, the approximation
made in ref.~\cite{Dvo19} is quite rough.

We should start with the specification of the initial condition for eq.~(\ref{eq:rhoIeq}),
$\rho_{\mathrm{I}}(0)=\rho(0)$. We study the evolution of astrophysical
neutrinos created in decays of charged pions~\cite{LeaPak95}. In
this case, one has that the fluxes of flavor neutrinos at a source
are $(F_{\nu_{e}}:F_{\nu_{\mu}}:F_{\nu_{\tau}})_{\mathrm{S}}=(1:2:0)$.
Thus $\rho_{\mathrm{I}}(0)=\text{diag}(1/3,2/3,0)$.

First, we study the GW background from coalescing supermassive BHs
(SMBHs). In this case, we can approximate $\Omega(f)$ by~\cite{Ros11}
\begin{equation}\label{eq:Omegaf}
  \Omega(f)=
  \begin{cases}
    \Omega_{0}, & \text{if}
    \quad
    f_{\mathrm{min}}<f<f_{\mathrm{max}},
    \\
    0, & \text{otherwise,}
\end{cases}
\end{equation}
where $\Omega_{0}\sim10^{-9}$, $f_{\mathrm{min}}\sim10^{-10}\,\text{Hz}$,
and $f_{\mathrm{max}}\sim10^{-1}\,\text{Hz}$. Note that this value
of $\Omega_{0}$ does not violate the constraint established in ref.~\cite{Abb19}.
Using eq.~(\ref{eq:Omegaf}), we get $g(t)$ in eq.~(\ref{eq:gspec})
in the explicit form,
\begin{align}\label{eq:gt}
  g(t)= & -\frac{2G\rho_{c}\Omega_{0}}{3\pi^{2}}
  \bigg\{
    \sin
    \left(
      2\pi f_{\mathrm{max}}t
    \right)
    \frac{2-(2\pi f_{\mathrm{max}}t)^{2}}{f_{\mathrm{max}}^{3}}-\sin
    \left(
      2\pi f_{\mathrm{min}}t
    \right)
    \frac{2-(2\pi f_{\mathrm{min}}t)^{2}}{f_{\mathrm{min}}^{3}}
    \nonumber
    \\
    & +
    2\pi t
    \left[
      \frac{\cos
      \left(
        2\pi f_{\mathrm{max}}t
      \right)}
      {f_{\mathrm{max}}^{2}}-\frac{\cos
      \left(
        2\pi f_{\mathrm{min}}t
      \right)}
      {f_{\mathrm{min}}^{2}}
    \right]
    \notag
    \\
    & +
    (2\pi t)^{3}
    \left[
      \mathrm{Ci}
      \left(
        2\pi f_{\mathrm{max}}t
      \right)-
      \mathrm{Ci}
      \left(
        2\pi f_{\mathrm{min}}t
      \right)
    \right]
  \bigg\},
\end{align}
where $\mathrm{Ci}(x)=\gamma+\ln x+\smallint_{0}^{x}\tfrac{\cos t-1}{t}\mathrm{d}t$
is the cosine integral and $\gamma\approx0.577$.

\begin{figure}
  \centering
  \subfigure[]
  {\label{1a}
  \includegraphics[scale=.38]{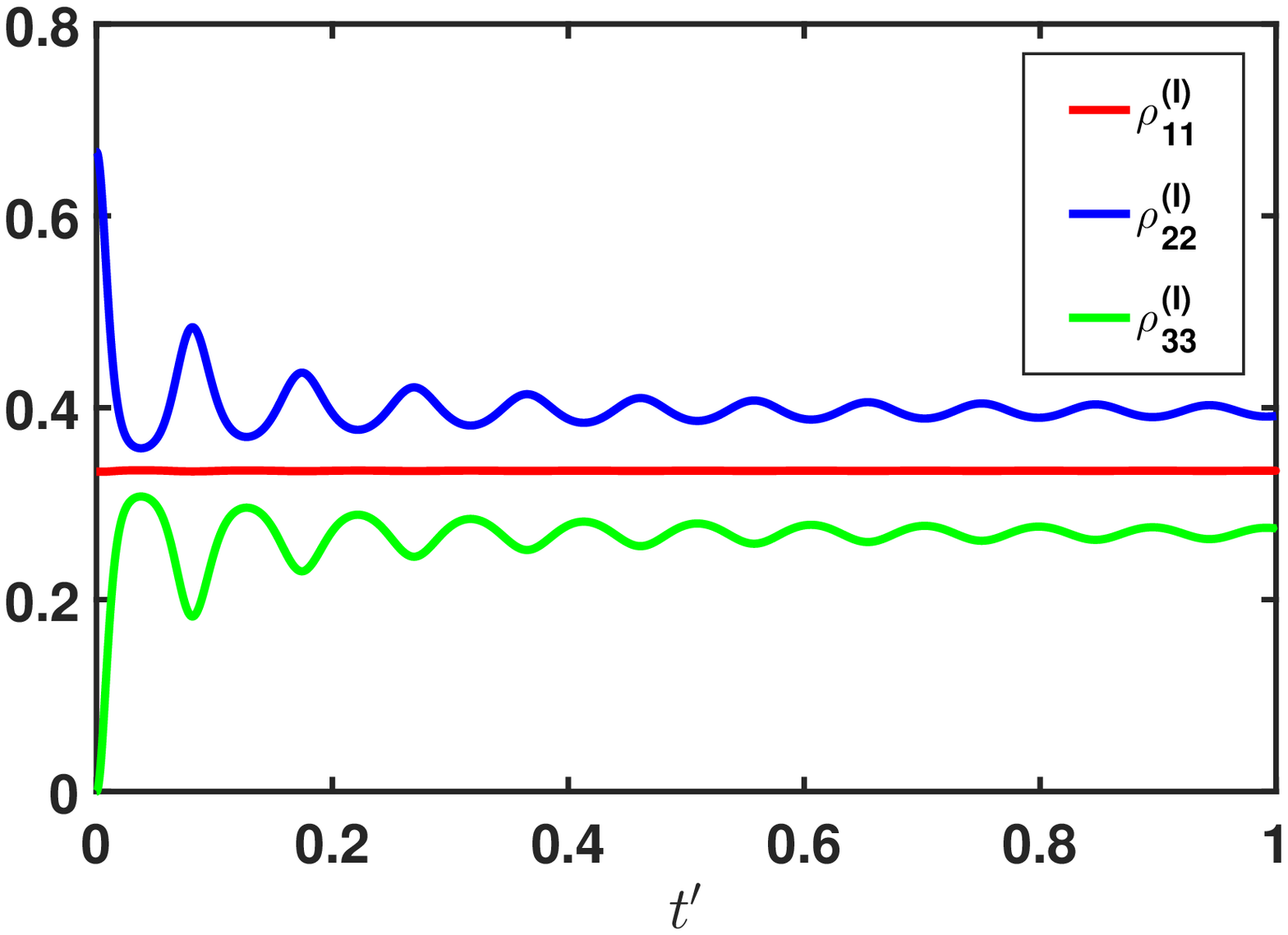}}
  \hskip-.6cm
  \subfigure[]
  {\label{1b}
  \includegraphics[scale=.38]{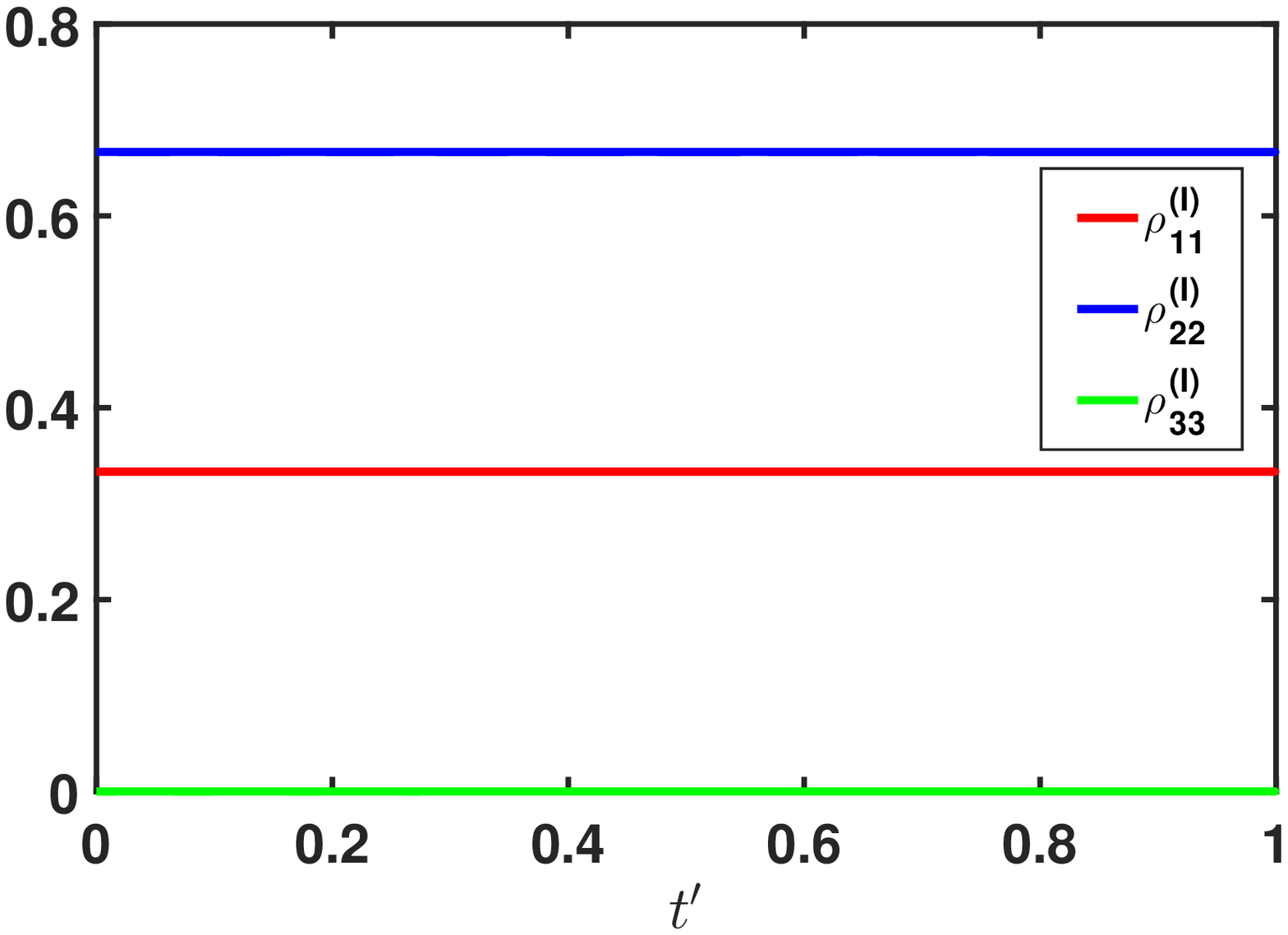}}
  \protect
  \caption{The diagonal elements of $\left\langle \rho_{\mathrm{I}}\right\rangle$
  based on the numerical solution of eq.~(\ref{eq:rhoIeq}) versus
  $t'=t/L$. Here we adopt the normal mass ordering with
  $\Delta m_{21}^{2}  =7.5\times10^{-5}\,\text{eV}^{2}$,
  $\Delta m_{31}^{2}=2.56\times10^{-3}\,\text{eV}^{2}$, $\theta_{12}=0.6$,
  $\theta_{23}=0.85$, $\theta_{13}=0.15$, and $\delta_{\mathrm{CP}}=3.77$  (see ref.~\cite{Sal20}).
  The neutrino beam propagation distance is
  $L=1\,\text{kpc}$. (a)~$E=10^{2}\,\text{MeV}$;
  and (b)~$E=10\,\text{GeV}$.\label{fig:normal}}
\end{figure}

\begin{figure}
  \centering
  \subfigure[]
  {\label{2a}
  \includegraphics[scale=.38]{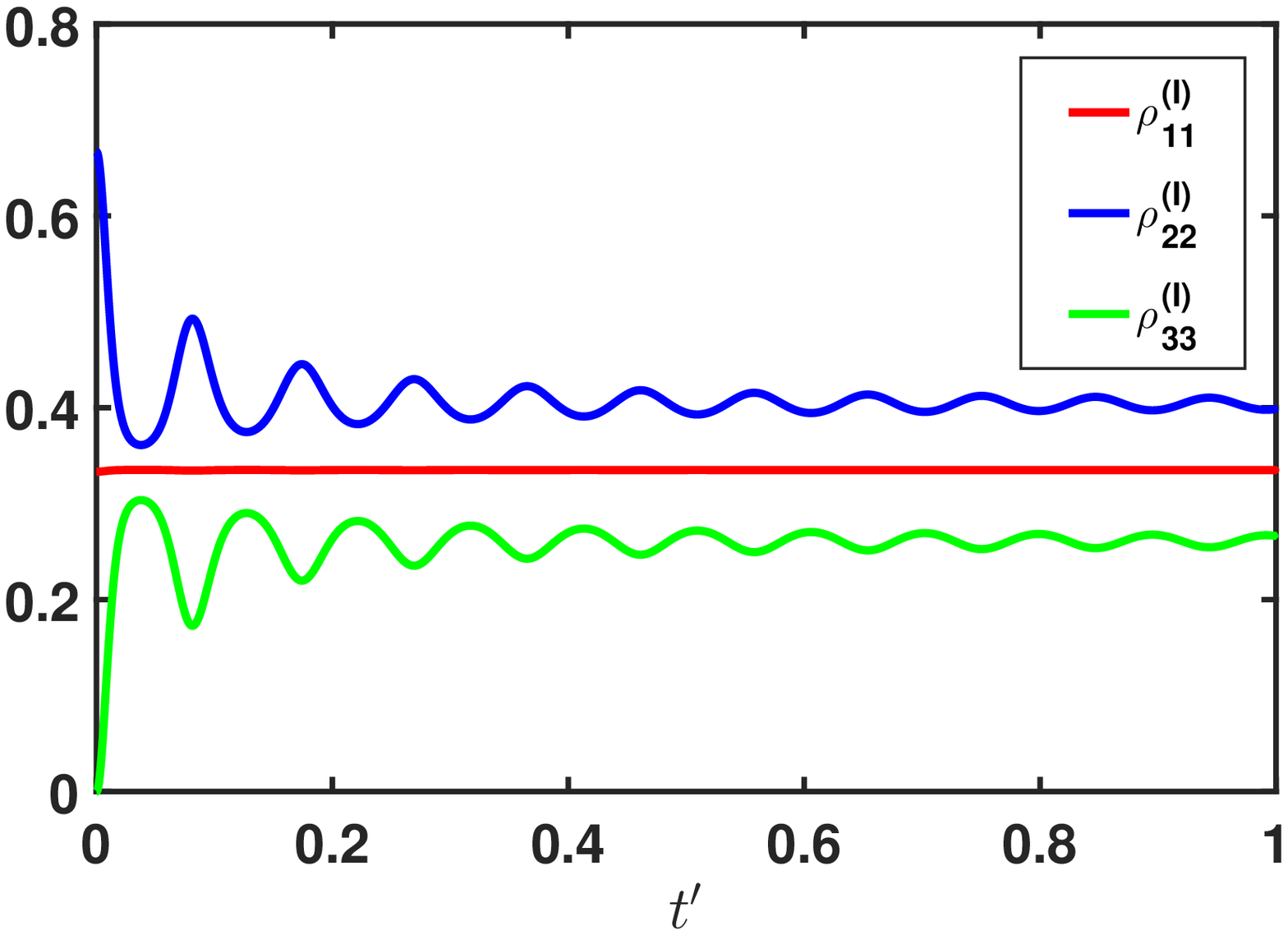}}
  \hskip-.6cm
  \subfigure[]
  {\label{2b}
  \includegraphics[scale=.38]{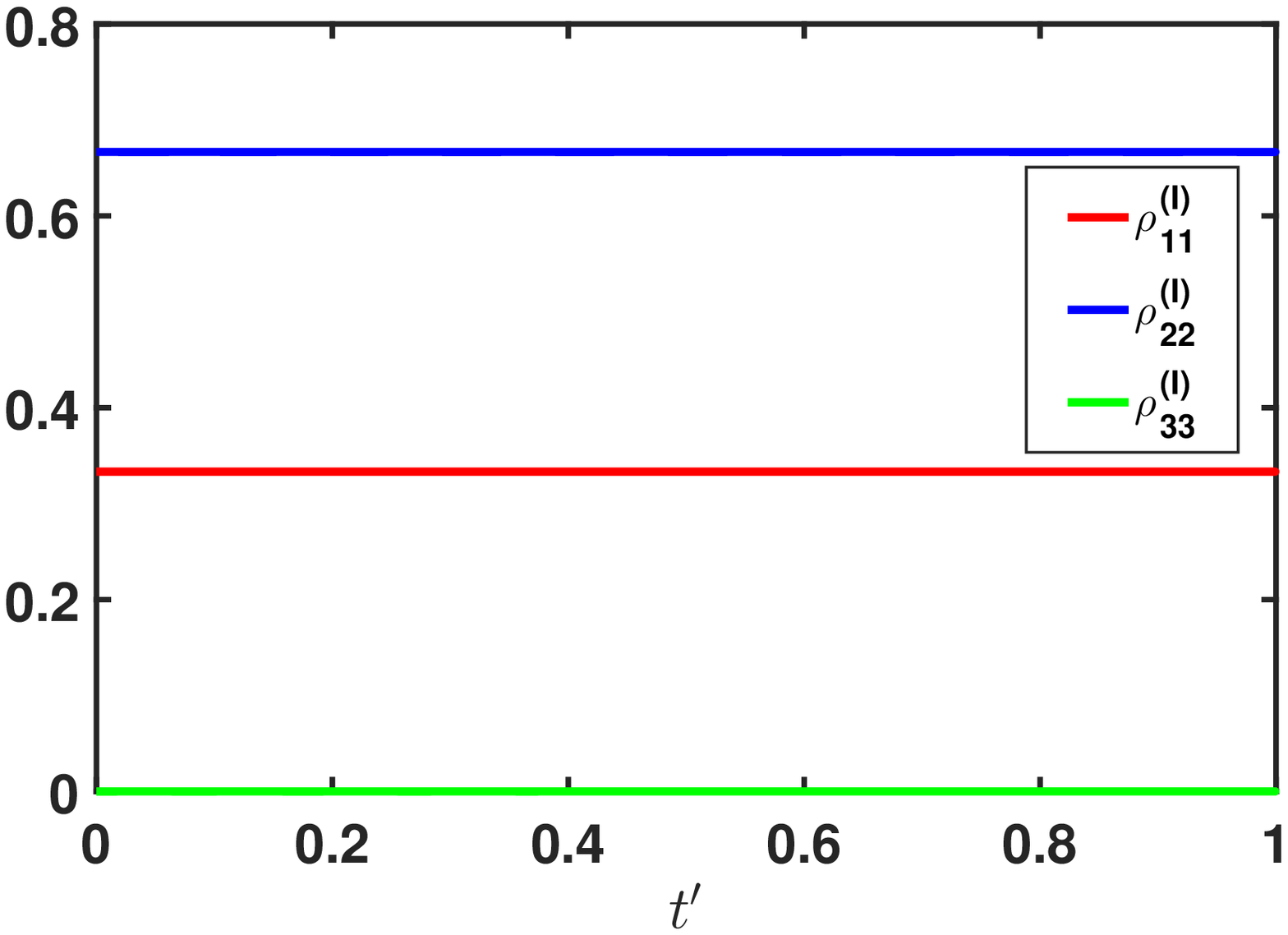}}
  \protect
  \caption{The same as in figure~\ref{fig:normal}, but for the inverted mass
  ordering with $\Delta m_{21}^{2}=7.5\times10^{-5}\,\text{eV}^{2}$,
  $\Delta m_{31}^{2}=-2.46\times10^{-3}\,\text{eV}^{2}$, $\theta_{12}=0.59$,
  $\theta_{23}=0.85$, $\theta_{13}=0.15$, and $\delta_{\mathrm{CP}}=4.84$.
  (a)~$E=10^{2}\,\text{MeV}$; and (b)~$E=10\,\text{GeV}$.  
  \label{fig:inverted}}
\end{figure}

In figures~\ref{fig:normal} and~\ref{fig:inverted}, we show the
numerical solution of eqs.~(\ref{eq:rhoIeq}) and~(\ref{eq:gt})
with $\left\langle A^{2}\right\rangle =\tfrac{3}{64}$, $\Omega_{0}$,
and $f_{\mathrm{min},\mathrm{max}}$ given above. The neutrino energy
is in the range $E=(10^{2} - 10^{4})\,\text{MeV}$ and the propagation
distance is $L=1\,\text{kpc}$. We study the cases of both the normal
and inverted orderings of the neutrino masses. The mass squared differences
and the mixing angles are taken from ref.~\cite{Sal20}.

One can see in figures~\ref{1b} and~\ref{2b}
that the contribution of GWs to neutrino oscillations is negligible
at high neutrino energies since $H_{m}^{(g)}\sim E^{-1}$ in eq.~(\ref{eq:gravHam}).
Indeed, one gets that $\left\langle \rho_{kk}^{(\mathrm{I})}\right\rangle (t)$
practically coincides with $\left\langle \rho_{kk}^{(\mathrm{I})}\right\rangle (0)=\text{diag}(1/3,2/3,0)$
at $E=10\,\text{GeV}$.

Comparing the results in figures~\ref{1a} and~\ref{2a} with the findings of ref.~\cite{Dvo19},
where analogous problem was studied, we establish the more significant
effect of GWs on the relaxation of the density matrix now. We find that $\langle \rho_\mathrm{I} \rangle$ reaches its asymptotic values at shorter
$L$ and greater $E$. This discrepancy can be explained by the underestimation
of the relaxation time $\tau$ in ref.~\cite{Dvo19}. In the present
work, we take into account the spectrum of stochastic GWs exactly.
Thus there is no need to approximate the correlator by a $\delta$-function.

\begin{figure}
  \centering
  \subfigure[]
  {\label{3a}
  \includegraphics[scale=.38]{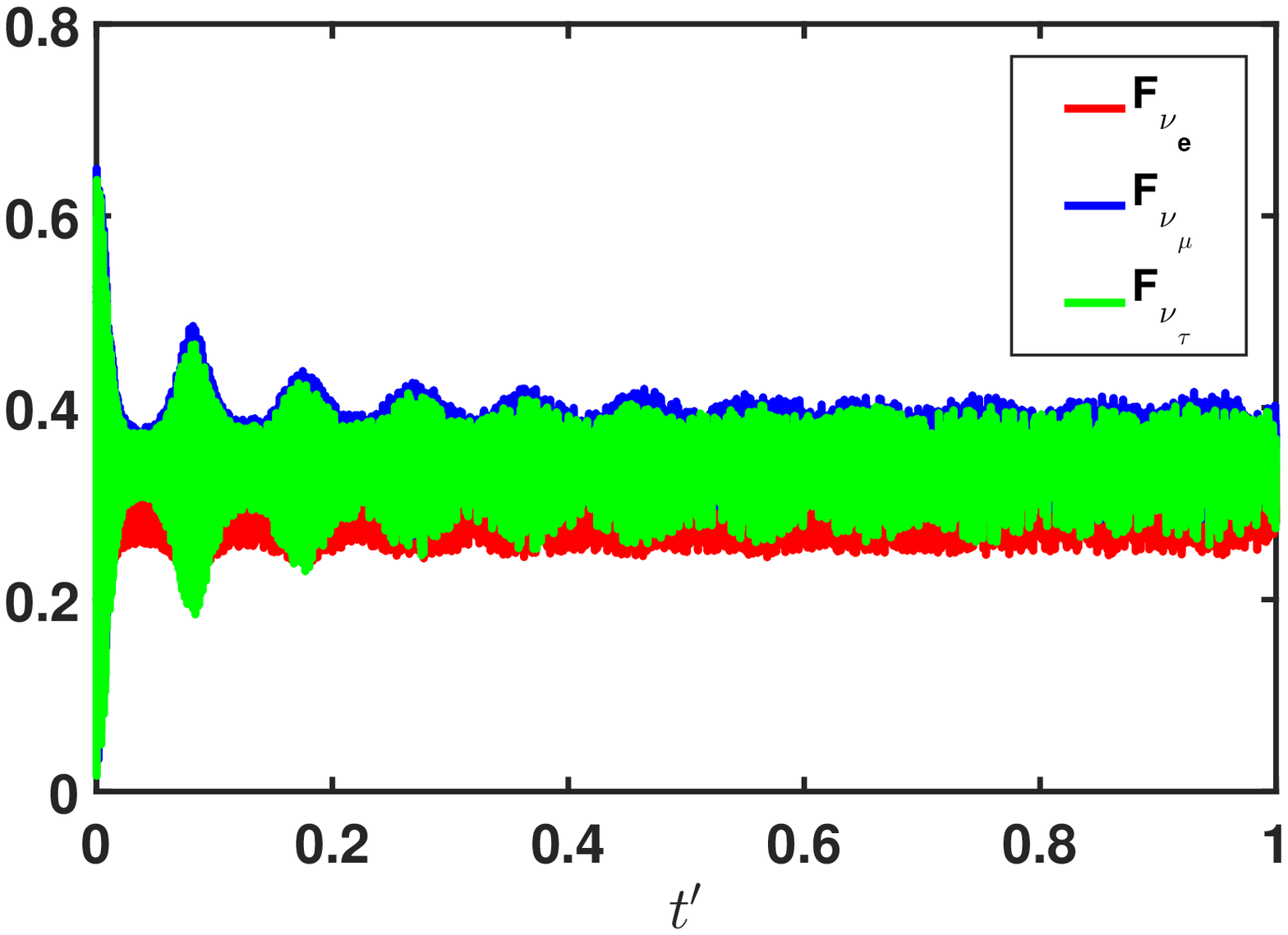}}
  \hskip-.6cm
  \subfigure[]
  {\label{3b}
  \includegraphics[scale=.38]{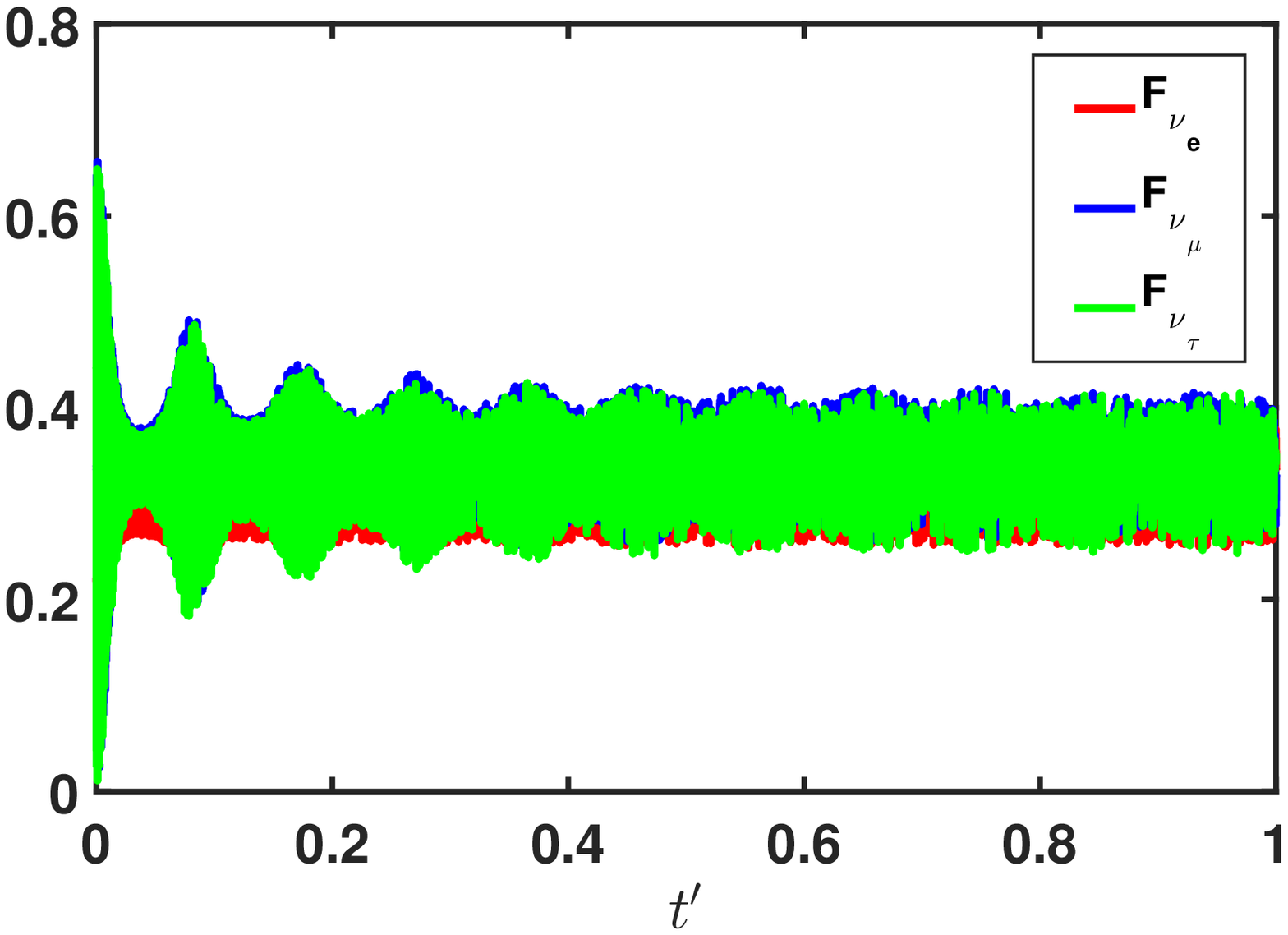}}
  \protect
  \caption{Fluxes of flavor neutrinos based on the numerical solution of   
  eq.~(\ref{eq:rhoIeq}) with $g(t)$ in eq.~(\ref{eq:gt}) for
  $E=10^{2}\,\text{MeV}$ and $L=1\,\text{kpc}$.
  (a) Normal mass ordering; and (b) inverted mass ordering.\label{fig:Fluxes}}
\end{figure}

The elements of $\left\langle \rho_{\mathrm{I}}\right\rangle $ are
not the measurable quantities. The fluxes of flavor neutrinos $F_{\nu_{\lambda}}$
are proportional to the diagonal elements of the total density matrix
$\left\langle \rho\right\rangle (t)=\exp(-\mathrm{i}H_{0}t)\left\langle \rho\right\rangle _{\mathrm{I}}(t)\exp(\mathrm{i}H_{0}t)$.
In figure~\ref{fig:Fluxes}, we show $F_{\nu_{\lambda}}$ for $L=1\,\text{kpc}$
and $E=10^{2}\,\text{MeV}$. We have demonstrated in figures~\ref{1b}
and~\ref{2b} that GWs does not contribute to the evolution
of the density matrix at $E=10\,\text{GeV}$. Thus we do not show the
fluxes for $E=10\,\text{GeV}$.

It is difficult to distinguish the fluxes for different flavors in
figure~\ref{fig:Fluxes} by sight because of the rapid vacuum oscillations.
However, if one averages the signal over the length of these oscillations,
one gets the fluxes in a detector $\left\langle F_{\nu_{\lambda}}\right\rangle_\oplus$
which turn out to be different for different flavors. This additional
averaging is equivalent to the situation when we study not only the
action of stochastic GWs on the neutrino beam, but also consider randomly
distributed neutrino sources.

We write down the values of $\left\langle F_{\nu_{\lambda}}\right\rangle_\oplus$ for $E=10\,\text{MeV}$ and $L=1\,\text{kpc}$ in the upper rows in tables~\ref{tab:normal} and~\ref{tab:inverted} for the normal and inverted mass orderings. Firstly, one can see that the cases of the
normal and inverted mass orderings are distinguishable. Secondly, $\left\langle F_{\nu_{\lambda}}\right\rangle_\oplus$ are not equal for different neutrino flavors. This feature is discussed in appendix~\ref{sec:FLUXES}. We remind that we adopt the following ratio of the fluxes in a source: $(1:2:0)_\mathrm{S}$.

\begin{table}
  \centering
  \begin{tabular}{|p{7cm}|p{1.5cm}|p{1.5cm}|p{1.5cm}|}
    \hline 
    {} &
    $\left\langle F_{\nu_{e}}\right\rangle_\oplus$ &
    $\left\langle F_{\nu_{\mu}}\right\rangle_\oplus$ &
    $\left\langle F_{\nu_{\tau}}\right\rangle_\oplus$
    \tabularnewline
    \hline 
    \hline 
    Accounting for stochastic GWs and averaging over the source position
    & 0.3038 & 0.3551 & 0.3411
    \tabularnewline
    \hline 
    Only averaging over the source position & 0.3029 & 0.3552 & 0.3419
    \tabularnewline
    \hline
    Relative contribution of stochastic GWs (absolute value, in \%)
    & 0.3 & 0.03 & 0.2
    \tabularnewline
    \hline 
  \end{tabular}
  \caption{Averaged fluxes of flavor neutrinos in a detector for the normal mass ordering for $E = 10^2\,\text{MeV}$. The upper row is based on figure~\ref{3a}. The middle row corresponds to eq.~\eqref{eq:PLaveraged}. The lower row represents the absolute value of the relative contribution of stochastic GWs.\label{tab:normal}}
\end{table}

\begin{table}
  \centering
  \begin{tabular}{|p{7cm}|p{1.5cm}|p{1.5cm}|p{1.5cm}|}
    \hline 
    {} &
    $\left\langle F_{\nu_{e}}\right\rangle_\oplus$ &
    $\left\langle F_{\nu_{\mu}}\right\rangle_\oplus$ &
    $\left\langle F_{\nu_{\tau}}\right\rangle_\oplus$
    \tabularnewline
    \hline 
    \hline 
    Accounting for stochastic GWs and averaging over the source position
    & 0.3190 & 0.3440 & 0.3370
    \tabularnewline
    \hline 
    Only averaging over the source position & 0.3182 & 0.3456 & 0.3362
    \tabularnewline
    \hline
    Relative contribution of stochastic GWs (absolute value, in \%)
    & 0.3 & 0.5 & 0.2
    \tabularnewline
    \hline 
  \end{tabular}
  \caption{The same as in table~\ref{tab:normal}, but for the inverted mass
  ordering. The upper row is based on figure~\ref{3b}.
%The middle row corresponds to eq.~\eqref{eq:PLaveraged}. The lower row shows the absolute value of the relative contribution of stochastic GWs.
\label{tab:inverted}}
\end{table}

One can evaluate the effect of stochastic GWs on the measured fluxes in a detector using the data in tables~\ref{tab:normal} and~\ref{tab:inverted}. For this purpose, in the middle rows in tables~\ref{tab:normal} and~\ref{tab:inverted}, we represent the fluxes in a detector in the case when only the averaging over the distance between a source and a detector is made, without accounting for the action of GWs. These values are based on eq.~\eqref{eq:PLaveraged} and also correspond to $(1:2:0)_\mathrm{S}$. The neutrino mixing angles for both the normal and inverted mass orderings are taken from ref.~\cite{Sal20}.

The difference between the upper and middle rows in tables~\ref{tab:normal} and~\ref{tab:inverted} is the measure of the contribution of stochastic GWs to the fluxes of flavor neutrinos. It is represented in the lower rows in tables~\ref{tab:normal} and~\ref{tab:inverted}. One can see that this contribution is below 1\%.

%The values of $\left\langle F_{\nu_{\lambda}}\right\rangle_\oplus $
%for different $E$ are given in tables~\ref{tab:normal} and~\ref{tab:inverted}.
%Here we take that $L=1\,\text{kpc}$.  Moreover there
%is a difference of the fluxes for relatively low, $E=10^{2}\,\text{MeV}$,
%and high, $E=10\,\text{GeV}$, energies. We remind that we established
%above that, at $E=10\,\text{GeV}$, stochastic GWs practically do
%not contribute to the relaxation of neutrino fluxes. Therefore, the
%difference of the fluxes in the first and the second rows in tables~\ref{tab:normal}
%and~\ref{tab:inverted} is owing to the neutrino interaction with
%stochastic GWs. One can see that this difference can be up to $2\,\%$.
%It is maximal for the inverted mass ordering.

We note that the above analysis is valid if the constraint in eq.~(\ref{eq:constr})
is fulfilled. To get the maximal left hand side in eq.~(\ref{eq:constr})
we take $\omega=\omega_{\mathrm{max}}=2\pi f_{\mathrm{max}}$, $E=10^{2}\,\text{MeV}$
and $L=1\,\text{kpc}$. We obtain that this condition is satisfied for such values.

Now we briefly consider another source of stochastic GWs, like merging
binary BHs of stellar masses. We can take that $\alpha=2/3$~\cite{Abb19}.
The spectral density has the form,
\begin{equation}\label{eq:OmegafBH}
  \Omega(f)=
  \begin{cases}
    \Omega_{\mathrm{min}}
    \left(
      \frac{f}{f_{\mathrm{min}}}
    \right)^{2/3}, &
    \text{if}
    \quad
    f_{\mathrm{min}}<f<f_{\mathrm{max}},\\
    0, & \text{otherwise,}
  \end{cases}
\end{equation}
where $f_{\mathrm{min}}=10^{-5}\,\text{Hz}$, $f_{\mathrm{max}}=10^{2}\,\text{Hz}$,
and $\Omega_{\mathrm{min}}=10^{-15}$~\cite{Ros11}. Unfortunately,
it is not possible to express the function $g(t)$ in eq.~(\ref{eq:gspec})
in the explicit form for $\Omega(f)$ in eq.~(\ref{eq:OmegafBH}).

We have solved eq.~(\ref{eq:rhoIeq}) for this case. The value of
$\left\langle \rho_{\mathrm{I}}\right\rangle (t)$ turns out to be
unchanged for $f_{\mathrm{min,max}}$ and $\Omega_{\mathrm{min}}$
given above for both normal and inverted mass orderings. Thus the
evolution of $\left\langle \rho_{\mathrm{I}}\right\rangle (t)$ is
similar to that shown in figures~\ref{1b} and~\ref{2b}.
We have checked $E$ down to $1\,\text{MeV}$ and $L$ up to $1\,\text{Gpc}$.

Thus, merging BHs with stellar masses as sources stochastic GWs do
not result in an effective relaxation of the fluxes of astrophysical neutrinos.
Such sources of GWs are more abundant than SMBHs studied above. Nevertheless
the typical frequencies of the spectrum of GWs emitted are much higher
that in case of SMBHs. It is the main reason why this kind of GWs does not
contribute to the relaxation of neutrino fluxes.

\section{Discussion\label{sec:DISC}}

In the present work, we have studied the evolution of three mixed
flavor neutrinos accounting for their interaction with stochastic
GWs with arbitrary spectrum. In section~\ref{sec:DENSMATR}, we have
derived eqs.~(\ref{eq:rhoIeq}) and~(\ref{eq:g}) for the density
matrix for flavor neutrinos. This equation generalizes the result
of ref.~\cite{Dvo19}, where the $\delta$-correlator of the GW amplitudes
was assumed.

In realistic situations, the spectral density of stochastic GWs is
a certain function of the frequency in a confined frequency range.
It leads to the correlator of amplitudes not necessarily proportional
to a $\delta$-function. Thus, the approximation made in ref.~\cite{Dvo19}
is quite rough.

In ref.~\cite{Dvo19}, the estimate of the correlation time $\tau$, or the typical frequency
of the spectrum $\tilde{f}\sim\tau^{-1}$, 
is the main source of the inexactitude in the description of the relaxation
of neutrino fluxes. Indeed, $g\sim\tilde{f}^{-3}$ in eq.~(\ref{eq:gspec})
and $\tilde{f}$ is in a quite broad range. Therefore, slightly changing
$\tilde{f}$, we get a significant variation of the relaxation length
of the neutrino fluxes.

In the present work, we have avoided this uncertainty. The lower part of
the spectrum $\Omega(f)$ turns out to give the main contribution
to the relaxation of the fluxes. It can explain much faster relaxation
of the fluxes compared to ref.~\cite{Dvo19}.

We have considered the application of our results for the evolution
of fluxes of astrophysical neutrinos in section~\ref{sec:APPL}. We
have studied merging binary BHs as sources of stochastic gravitational
waves. We have considered two cases: SMBHs and BHs with stellar masses.
In the case of SMBHs, the relaxation distance should be $L\gtrsim1\,\text{kpc}$ for the fluxes to reach their asymptotic values.
The relaxation of fluxes is sizable for neutrinos with energies $E  \lesssim 10^{2}\,\text{MeV}$. Thus, the effect of the relaxation of the neutrino
fluxes can be important for supernova (SN) neutrinos in our galaxy, with $L_\mathrm{MW} = 32\,\text{kpc}$, since the typical energy of such neutrinos $E_\mathrm{SN} \sim 10\,\text{MeV}$.

A strong evidence for the existence of the stochastic GWs background, generated by SMBHs, is reported in ref.~\cite{Arz20} recently. The typical frequency of the observed signal is $\sim 10^{-8}\,\text{s}^{-1}$. Such a frequency is within the range used in eq.~\eqref{eq:Omegaf}.

The fluxes of SN neutrinos were reported in ref.~\cite{Ker07} to
be modified by the mixing between active and hypothetical sterile
neutrinos. We have demonstrated that the interaction of active neutrinos
with stochastic GWs background with realistic characteristics can
also slightly modify the observed SN neutrino fluxes.
Perhaps, the predicted effect can be observed by the existing~\cite{Kul17}
or future~\cite{Mig18,Fis18} neutrino telescopes.

Merging BHs with stellar masses are more abundant than coalescing
SMBHs. However the lower frequency of their spectra is much higher
than that of SMBHs; cf. eqs.~(\ref{eq:Omegaf}) and~(\ref{eq:OmegafBH}).
Thus the effect of the relaxation of neutrino fluxes is smaller in this
case. This fact has been confirmed by our numerical simulations.

We have revealed that the neutrino interaction with stochastic GWs
results in the relaxation of neutrino fluxes. However, it is not the
only random factor affecting the observed fluxes of astrophysical
neutrinos. In a realistic situation, a neutrino telescope detects
particles emitted by multiple randomly distributed sources.

In section~\ref{sec:APPL}, we have accounted for this factor by considering
oscillations in vacuum and averaging the observed fluxes over the distance between a detector and sources. The corresponding fluxes are derived in appendix~\ref{sec:FLUXES}; cf. eq.~\eqref{eq:PLaveraged}. Comparing the case when stochastic GWs and the distance averaging are accounted for with the situation when only the distance averaging is made, one can extract the contribution of stochastic GWs to the measured neutrino fluxes; cf. tables~\ref{tab:normal} and~\ref{tab:inverted}. This contribution turns out to be below 1\% level.

Despite the obtained contribution of GWs is small, one can express a restrained hope that a development of experimental techniques will allow one to explore stochastic GWs through a precise measurement
of astrophysical neutrino fluxes rather than using direct methods
described in ref.~\cite{RomCor17}.

Accounting for the interaction with stochastic GWs and averaging over
the positions of randomly distributed neutrino sources, we have obtained
that the fluxes at a detector are not equal: $\left(F_{\nu_{e}}:F_{\nu_{e}}:F_{\nu_{e}}\right)_{\oplus}\neq(1:1:1)$. We have demonstrated in appendix~\ref{sec:FLUXES} that the situation $\left(F_{\nu_{e}}:F_{\nu_{e}}:F_{\nu_{e}}\right)_{\oplus} = (1:1:1)$ is possible only when the initial ratio of fluxes is $(1:1:1)_\mathrm{S}$. 
This problem of the flavor content of astrophysical neutrinos was
studied in ref.~\cite{NunPanZuk16}. The fluxes of ultra high energy
neutrinos at a detector were found in ref.~\cite{NunPanZuk16} to
depend on the channel of the production of these particles.

In the present work, we have demonstrated that the interaction of
astrophysical neutrinos with stochastic GWs can result in the deviation
of the flavor ratio in a detector from the value $(1:1:1)$. The predicted
fluxes are not excluded by the recent observation of ultra high energy
astrophysical neutrinos reported in ref.~\cite{Aar15}. Plans to
improve the sensitivity in the determination of the flavor ratio of astrophysical
neutrinos are outlined in ref.~\cite{Aar20}.

In a realistic situation, one registers the total flux of astrophysical neutrinos with a terrestrial detector, without any information about neutrino sources. Moreover, a specific neutrino source can have a certain energy spectrum. Thus, one has to average the final fluxes over the initial ones accounting for their nontrivial energy spectra.\footnote{In the present work, we adopt the simplest situation of a monocromatic neutrino source.} The same procedure should be made when neutrino oscillations are considered in the presence of stochastic GWs. To evaluate the contribution of GWs, one has to compare these two cases, as it was made in section~\ref{sec:APPL}. This contribution is expected to be quite small. However, a careful separate analysis is required to get a quantitative result.

In summary, we have studied the evolution of fluxes of astrophysical
neutrinos interacting with stochastic GWs having an arbitrary energy
spectrum emitted by randomly distributed realistic binary BHs. The
consideration of the nontrivial spectrum allowed us to significantly
reduce the relaxation distance traveled by a neutrino beam for different
flavors to reach the asymptotic values. We could also increase the
neutrino energy, for which the relaxation becomes sizable. Now the
predicted effect can be potentially observed even for SN neutrinos propagating
within our Galaxy.

\acknowledgments

This work is performed within the government assignment of IZMIRAN.

\appendix

\section{Fluxes of neutrinos emitted by random sources\label{sec:FLUXES}}

In this appendix, we derive the expressions for the probabilities to detect a certain flavor of neutrinos produced by multiple, randomly distributed sources accounting for only flavor oscillations in vacuum.

We suppose that a beam of flavor neutrinos is emitted with the probabilities $P_\lambda(0) = |\nu_\lambda(0)|^2$, where $\lambda = e,\mu,\tau$ and $|\nu_\lambda(0)\rangle$ is the wave function of flavor neutrinos at a source. This neutrino beam is observed in a detector, which is at the distance $L\approx t$ from a source. We suppose that these neutrinos propagate in vacuum.

Using eqs.~\eqref{eq:nupsi} and~\eqref{eq:Hf}, where we keep only $H_0$ in $H_f$, we get the wave function of flavor neutrinos in a detector,
\begin{equation}\label{eq:nunu0}
  |\nu_\lambda(L)\rangle = \sum_{a\sigma}
  U_{\lambda a} U_{\sigma a}^* e^{-\mathrm{i}E_a L} |\nu_\sigma(0)\rangle,
\end{equation}
where $E_a = \sqrt{p^2 + m_a^2}$ is the energy of the mass eigenstate and $(U_{\lambda a})$ are the components of the mixing matrix $U$.

Using eq.~\eqref{eq:nunu0} and ref.~\cite[pgs.~247--252]{GonMalSch16}, we get the probabilities to observe a cetrain flavor $\lambda$ in this neutrino beam,
\begin{align}\label{eq:PL}
  P_\lambda(L) = & P_\lambda(0) + 2 \sum_{\sigma} P_\sigma(0)
  \Bigg[
    \sum_{a>b} \text{Im}
    \left(
      U_{\lambda a} U_{\lambda b}^* U_{\sigma b} U_{\sigma a}^*
    \right)
    \sin
    \left(
      \frac{\Delta m^2_{ab}}{2E} L
    \right) 
    \notag
    \\
    & -
    2\sum_{a>b} \text{Re}
    \left(
      U_{\lambda a} U_{\lambda b}^* U_{\sigma b} U_{\sigma a}^*
    \right)
    \sin^2
    \left(
      \frac{\Delta m^2_{ab}}{4E} L
    \right)
  \Bigg].
\end{align}
To derive eq.~\eqref{eq:PL} we take that $\langle \nu_\sigma(0) | \nu_\kappa (0) \rangle = \delta_{\sigma\kappa} P_\sigma(0)$.

Now we suppose that there are multiple neutrino sources with the random distribution with respect to a detector. Such a detector will register a neutrino background averaged over the source position. Therefore we account for the identities,
\begin{equation}\label{eq:averageL}
  \left\langle
    \sin
    \left(
      \frac{\Delta m^2_{ab}}{2E} L
    \right)
  \right\rangle = 0,
  \quad
  \left\langle
    \sin^2
    \left(
      \frac{\Delta m^2_{ab}}{4E} L
    \right)
  \right\rangle = 
  \frac{1}{2}.
\end{equation}
The mechanism of the neutrino emission is supposed to be the same in all the sources, i.e. $P_\sigma(0)$ is unchanged in the averaging in eq.~\eqref{eq:averageL}. Finally, we get the averaged probabilities, or the fluxes, of flavor neutrinos in a detector,
\begin{equation}\label{eq:PLaveraged}
  \langle P_\lambda(L) \rangle = P_\lambda(0) - 2 \sum_{\sigma} P_\sigma(0)
  \sum_{a>b} \text{Re}
  \left(
    U_{\lambda a} U_{\lambda b}^* U_{\sigma b} U_{\sigma a}^*
  \right),
\end{equation}
which generalizes the result in ref.~\cite{Dvo19}, which was obtained in the two flavors approximation.

The averaged fluxes in a detector $\langle F_{\nu_\lambda} \rangle_\oplus \sim \langle P_\lambda(L) \rangle$ can be in the ratio $(1:1:1)_\oplus$, suggested in ref.~\cite{LeaPak95}, only if the fluxes in a source $(F_{\nu_\lambda})_\mathrm{S} \sim P_\lambda(0)$ are equal for any neutrino flavor, i.e. $P_\lambda(0) = 1/3$ for $\lambda = e,\mu\,\tau$, or $(1:1:1)_\mathrm{S}$. In general case, including the situation $(1:2:0)_\mathrm{S}$ in our work, $\left( \langle F_{\nu_e} \rangle : \langle F_{\nu_\mu} \rangle : \langle F_{\nu_\tau} \rangle \right)_\oplus \neq (1:1:1)$. As results from eq.~\eqref{eq:PLaveraged}, this ratio depends on $(F_{\nu_\lambda})_\mathrm{S}$, or $P_\lambda(0)$, and the mixing angles.

\end{document}